\documentclass[preprint,12pt]{elsarticle}
\usepackage{graphicx}
\usepackage{amsmath}
\usepackage{amssymb}
\biboptions{numbers,sort&compress}

\makeatletter
\def\ps@pprintTitle{%
 \let\@oddhead\@empty
 \let\@evenhead\@empty
 \def\@oddfoot{}%
 \let\@evenfoot\@oddfoot}
\makeatother

\usepackage[left=2 cm,right=2cm,top=2cm,bottom=2cm]{geometry}
\numberwithin{equation}{section}
\numberwithin{figure}{section}

\begin{document}

\begin{frontmatter}
\title{Q-lumps on a domain wall with a spin-orbit interaction}
\author[label1,label2]{N. Blyankinshtein}

\address[label1]{Institute for Theoretical and Experimental Physics, 117218, Moscow, Russia}
\address[label2]{Moscow Institute of Physics and Technology, 141700, Dolgoprudny, Russia}

\ead{blyankinshtein@itep.ru}

\begin{abstract}
The nonlinear O(3)~$\sigma$ model in (2+1) dimensions with an additional potential term admits solutions called Q-lumps, having both topological and Noether charges. We consider in 3+1-dimensional spacetime the theory with Q-lumps on a domain wall in the presence of spin-orbit interaction in the bulk and find interaction effects for a two-particle solution through perturbation theory and adiabatic approximation.
\end{abstract}

\begin{keyword}
Q-lumps \sep O(3) $\sigma$--model \sep domain wall \sep spin-orbit interaction

\end{keyword}
\end{frontmatter}

\section{Introduction}

Topological defects occur in many topics of field theory, cosmology, and condensed matter physics. There is a particular interest to study models with additional non-Abelian internal degrees of freedom localized on solitons (domain walls, strings, monopoles, etc.); see Refs. \cite{HananyTong,Konishi, ShifmanYung2004,ShifmanTallaritaYung,ShifmanKurianovich,ANOModuli,MoreonANO,Moninetal,onDefects}. The next possible step is to consider theories containing solitons on solitons (as in Ref. \cite{Nitta}). In this paper, we study Q-lumps on the domain wall (see a similar construction in Ref. \cite{Nitta1210.2233}, a full numerical solution in Ref. \cite{Nitta1403.1245}, and a generalization to higher dimensions in Ref. \cite{Nitta1211.4916}).

Q-lumps are Q-ball-type solutions of the nonlinear O(3)~$\sigma$ model with an additional potential term of some special form. They were discovered by R.~Leese \cite{Leese} and have nontrivial topological charge in addition to the nonzero conserved Noether charges. Properties of such configurations differ significantly from those of a pure $\sigma$-model solution even when the coupling constant is small, e.g., in the case of small perturbation to the action, which may be important from a physical point of view as every physical system is unlikely to contain no perturbation at all. The main result of R.~Leese was that he first managed to obtain explicit solutions to the O(3)~$\sigma$ model with a particular potential term and to investigate the stability and scattering effects of them, and later the mechanism of building of such configurations was generalized to arbitrary $\sigma$ models by E.~Abraham \cite{Abraham}.

One special feature of Q-lumps is that the model admits stationary many-soliton solutions that can be interpreted as noninteracting particles (relevant to initially motionless configuration). In this paper, we investigate whether and what interaction appears if one adds a spin-orbit interaction term in the bulk (as in Refs. \cite{ShifmanKurianovich},\cite{ANOModuli}, and \cite{Moninetal}).

The main body of the paper looks at the interaction of two Q-lumps on a domain wall. Section~2 presents the model we deal with and introduces all the objects (the wall, Q-lumps, and spin-orbit interaction term) in detail. Section~3 investigates interaction effects perturbatively and through adiabatic approximation; finally, effects found are outlined in Sec.~4.

\section{Model}

\subsection{Action}

The theory considered includes scalar fields $\phi\in\Re$ and ${\chi_i\in\Re,~i=1,2,3}$ in 3+1-dimensional Minkowski spacetime with metric signature $(+,-,-,-)$ and admits a domain wall with additional non-Abelian internal degrees of freedom localized on it. The Lagrangian of the theory is
\begin{align}
&\mathcal{L}=\mathcal{L}_{\phi}+\mathcal{L}_{\chi},\label{larg0} \\
&\mathcal{L}_\phi = \frac{1}{2} \partial_\mu \phi\partial^\mu \phi-\lambda(\phi^2-v^2)^2,\\
&\mathcal{L}_\chi = \frac{1}{2} \partial_\mu \chi_i \partial^\mu \chi_i-\gamma((\phi^2-\mu^2)\chi_i \chi_i+\beta (\chi_i \chi_i)^2)-
\frac{1}{2} \alpha^2(\chi_i \chi_i-\chi_3\chi_3).
\end{align}
We introduce space coordinates named $(x,y,x_3)$ putting designation $z$ aside for a complex coordinate $z=x+i y$ on the $(x,y)$ plane. To build a domain wall, let us look for solutions with separated variables $\phi=\phi(x_3)$, $\chi_i=\chi(x_3) S_i(t,x,y)$, where $S_i S_i = 1$,
and fields $S_i$ correspond to O(3)~$\sigma$ model. Thus, the Lagrangian takes the form
\begin{align}
\mathcal{L}=\mathcal{L}_{\phi}+\frac{1}{2}\chi^2 \partial_p S_i \partial^p S_i-\frac{1}{2}(\chi')^2 -\gamma((\phi^2-\mu^2)\chi^2+\beta \chi^4)
-\frac{1}{2} \alpha^2\chi^2(1-S_3^2)=\nonumber \\
= \mathcal{L}_{wall}+2 \chi^2 \mathcal{L}_{lump},
\end{align}
where $p=0,1,2$; differentiation over $x_3$ is denoted by a prime and
\begin{align}
\mathcal{L}_{lump}=\frac{1}{4}\partial_p S_i \partial^p S_i-\frac{1}{4} \alpha^2(1-S_3^2). \label{Lagr}
\end{align}
For convenience, we introduce complex scalar fields $u$, $u^*$ instead of $S_i$ through stereographic projection, $u=\frac{S_1+i S_2}{1-S_3}$; then,
\begin{align}
\mathcal{L}_{lump}= \frac{\partial_p u \partial^p u^*}{(1+u u^*)^2}-\frac{\alpha^2 u u^*}{(1+u u^*)^2}.
\end{align}
The motion equations derived from action (\ref{larg0}) are
\begin{align}
\phi'' = 4 \lambda (\phi ^2-v^2) \phi+2\gamma \chi^2 \phi, \label{phieq}\\
\chi'' - 2 \gamma (\phi ^2-\mu^2) \chi - 4 \beta \gamma \chi ^3 + 4 \chi \mathcal{L}_{lump} = 0,\\
\partial_p \partial ^p u-\frac{2u^*\partial_p u \partial^p u}{1+u u^*}+\frac{\alpha^2 u (1-u u^*)}{1+u u^*} =0.
\end{align}

\subsection{Q-lumps}

The model with Lagrangian (\ref{Lagr}) in three-dimensional spacetime was studied in detail by Robert Leese in Ref. \cite{Leese}: he managed to find an explicit form of solitonic solutions having both topological and Noether charges related, respectively, to homotopic group ${\pi_2(S^2)=\mathbb{Z}}$ and to internal rotation with constant velocity around $S_3$ [one can notice that potential term breaks O(3) symmetry of the model to O(2)].
These solitons were named "Q-lumps" (by analogy with S.~Coleman's "Q-balls" \cite{Coleman}). They are not forbidden by Derrick's theorem (e.g., Ref. \cite{Derrick}) as they are not static and are stabilized by charge. Let us write them and their properties explicitly.

Topological and Noether charges and the total energy of configuration look like
\begin{align}
&N=\frac{1}{4\pi}\int \overrightarrow{S} \centerdot [\partial_x \overrightarrow{S} \times \partial_y \overrightarrow{S}]d^2x=\frac{i}{2 \pi}\int \frac{\partial_x u^*\partial_y u-\partial_x u\partial_y u^*}{(1+u u^*)^2} d^2x, \\
&Q = \int \frac{1}{2}(S_2 \partial_t S_1 - S_1 \partial_t S_2) d^2x=i \int \frac{u^* \partial_t u - u\partial_t u^*}{(1+u u^*)^2} d^2x,\\
&\mathcal{E}_{lump}=\int \frac{\partial_t u \partial_t u^*+ \partial_i u \partial_i u^*+\alpha^2 u u^* }{(1+u u^*)^2} d^2x
\end{align}
and are linked by the Bogomolny bound
\begin{align}
\mathcal{E}_{lump} \ge 2 \pi |N|+|\alpha Q|. \label{bound}
\end{align}
Thus, configurations saturating the Bogomolny bound have minimal energy among solutions with given charges and are classically stable because of charge conservation. Conditions of saturations are
\begin{align}
\partial_i u \pm i \varepsilon_{ij}\partial_j u = 0 \qquad \mbox{and} \qquad \partial_t u \pm i \alpha u = 0, \label{condition}
\end{align}
and that immediately gives an explicit form of the solutions,
\begin{align}
u(t,x,y)=u_0(x,y)\operatorname{e}^{\pm i \alpha t}, \qquad u_0(x,y)=u_0(x\pm i y);
\end{align}
here, the function $u_0(z = x + i y)$ must be (anti)rational for energy to be finite. In that case, the degree of the function gives topological charge $N$ of the configuration, whereas the Noether charge $Q$ takes a finite value when $|N| \geq 2 $.

For instance, the ansatz
${u(t,z=x+i y)=\left(\frac{\lambda}{z}\right)^k\operatorname{e}^{i \alpha t}}$ corresponds to a radially symmetric soliton of topological charge $k$, while ${u(t,z)=\frac{\beta z +\gamma}{z^2+\delta z +\epsilon}\operatorname{e}^{i \alpha t}}$ includes all solutions with topological charge 2.
Configurations having several distant poles could be interpreted as many-particle solutions and one can study scattering processes on moduli space in the limit of low energies, e.g., Refs. \cite{monopoles} and \cite{vortices} considering scattering of monopoles and vortices and in particular Refs. \cite{leese1990} and \cite{ward}
about solitons of the O(3)~$\sigma$ model.

\subsection{Domain wall}

\begin{figure}[h]
\center
\includegraphics[scale=0.63]{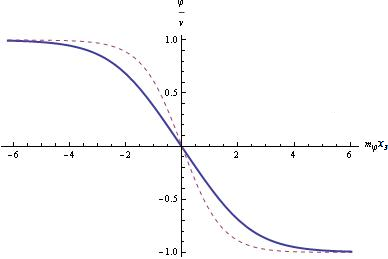}
\caption{Numerical solution for $\phi(x_3)$}
\label{pic1}
\end{figure}

\begin{figure}[h]
\center
\includegraphics[scale=0.63]{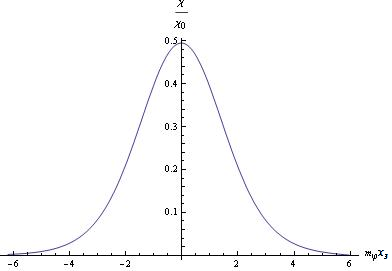}
\caption{Numerical solution for $\chi(x_3)$}
\label{pic2}
\end{figure}

The term $\mathcal{L}_{wall}=-\frac{1}{2}(\phi')^2-\lambda(\phi^2-v^2)^2-\frac{1}{2}(\chi')^2 -\gamma((\phi^2-\mu^2)\chi^2+\beta \chi^4)$ allows one to construct a static domain wall of fields $\phi(x_3)$ and $\chi(x_3)$. First, one can see that in the case of $\chi(x_3)=0$ the motion equation (\ref{phieq}) for the field $\phi(x_3)$ has ordinary kink solution $\phi(x_3)=-v \tanh{\frac{m_\phi}{2}\left(x_3-{x_3}_0\right)}$, but such a configuration is unstable (\cite{ShifmanKurianovich}).
The stable one has nonzero expectation value $\sqrt{\frac{\mu^2}{2 \beta}}$ inside the domain wall (i.e., around $x_3={x_3}_0$). Profiles of functions $\phi(x_3)$ and $\chi(x_3)$ shown in Figs.~\ref{pic1} and \ref{pic2} were derived numerically in the same paper \cite{ShifmanKurianovich} for some choice of parameters
$v,\lambda,\mu,\gamma,\beta $  of Lagrangian $\mathcal{L}_{wall}$.


Fields $S_i$ like field $\chi(x_3)$ become also localized on the wall around plane $x_3={x_3}_0$ and in this way describe effectively two-dimensional (to be more precise, 2+1-dimensional) theory on the domain wall.

\subsection{Spin-orbit interaction}

Now, we add spin-orbit interaction in the bulk (for more detail about its origin, see, for example, Ref. \cite{He}). It breaks Lorentz invariance of the Lagrangian,
leads to entanglement between fields $\chi_i$, and coordinates and is effectively described by the term
\begin{align}
\mathcal{L}_{so}=-\varepsilon (\partial_i \chi_i)^2
\end{align}
or, in the case of separated variables, $\chi_i=\chi(x_3) S_i(t,x,y)$,
\begin{align}
\mathcal{L}_{so}=-\varepsilon (\chi'^2(\frac{uu^*-1}{1+uu^*})^2+2\chi' \frac{uu^*-1}{1+uu^*}\chi \partial_k S_k +\chi^2 (\partial_k S_k)^2), \label{so}
\end{align}
where
\begin{align}
\partial_k S_k = \frac {1}{(1+uu^*)^2} \left( \partial_x u(1-{u^*}^2)+ \partial_x u^* (1-u^2) - i \partial_y u (1+{u^*}^2)+\right.\nonumber\\
\left.+i \partial_y u^* (1+u^2)\right).
\end{align}
To obtain the effective action for the theory on the wall in the presence of this term, one should integrate (\ref{larg0})~+~(\ref{so}) over $x_3$. Then, in 2+1 dimensions, we get
\begin{align}
\mathcal{L}_{eff} = A\left(\frac{\partial_p u \partial^p u^*}{(1+u u^*)^2}-\frac{\alpha^2 u u^*}{(1+u u^*)^2}\right)-\nonumber\\
-\varepsilon \left( B(\frac{uu^*-1}{1+uu^*})^2+C\frac{uu^*-1}{1+uu^*} \partial_k S_k + D (\partial_k S_k)^2 \right),
\end{align}
where constant values
\begin{align}
A=\int 2 \chi^2 dx_3,\\
B=\int (\chi')^2 dx_3,\\
C= \int 2 \chi' \chi dx_3,\\
D=\int \chi^2 dx_3=\frac{1}{2}A,
\end{align}
are introduced and in the case of the symmetric wall [function $\chi(x_3)$]
\begin{align}
C=0.
\end{align}
After division by a constant,
\begin{align}
&\mathcal{L}_{eff} = \frac{\partial_p u \partial^p u^*}{(1+u u^*)^2}-\frac{\alpha^2 u u^*}{(1+u u^*)^2}-\varepsilon \left( \frac{B}{A}(\frac{uu^*-1}{1+uu^*})^2+ \frac{1}{2} (\partial_k S_k)^2 \right)= \nonumber\\
&=\frac{1}{4}\partial_p S_i \partial^p S_i-\frac{1}{4} \alpha^2(1-S_3^2)-\varepsilon \left( \frac{B}{A} S_3^2 +\frac{1}{2} (\partial_k S_k)^2 \right) \label{efac}
\end{align}
and up to a constant term,
\begin{align}
\mathcal{L}_{eff} = \frac{1}{4}\partial_p S_i \partial^p S_i-\frac{1}{4} \alpha'^2(1-S_3^2)-\frac{\varepsilon}{2} (\partial_k S_k)^2 = \frac{\partial_p u \partial^p u^*}{(1+u u^*)^2}-\frac{\alpha'^2 u u^*}{(1+u u^*)^2}-\nonumber\\
-\frac{\varepsilon}{2} \frac{( \partial_x u(1-{u^*}^2)+ \partial_x u^* (1-u^2) -i \partial_y u (1+{u^*}^2)+i \partial_y u^* (1+u^2))^2}{(1+uu^*)^4} \label{eff}.
\end{align}
Thereby, of the three terms in Eq. (\ref{so}) the first one leads to the correction of $\alpha$,
\begin{align}
\alpha \rightarrow \alpha'= \sqrt{\alpha^2-4\varepsilon \frac{B}{A}},
\end{align}
the second one vanishes due to the symmetry, and only the third one can significantly influence the system.

\section{Interaction}

To find interaction effects, let us consider a two-particle solution of nonperturbated theory (without spin-orbit interaction). Two-particle configurations correspond to solutions of topological charge 2:
\begin{align}
u_0(t,z=x+iy)=\frac{\gamma}{z^2+\epsilon}\operatorname{e}^{i \alpha' t}=
\frac{\gamma}{i \sqrt{\epsilon}}\left( \frac{1}{z-i \sqrt{\epsilon}}-\frac{1}{z+i \sqrt{\epsilon}} \right) \operatorname{e}^{i \alpha' t}. \label{nonper}
\end{align}
Here, complex parameters $\gamma$ and $\epsilon$ give the size of particles $\left( \frac{| \gamma| }{2 \sqrt{|\epsilon|}} \right)$ and the distance between them
$\left( 2 \sqrt{|\epsilon|}\right)$.
\subsection{Perturbation theory}
Here, we look for a correction to ansatz (\ref{nonper}) by perturbation theory, assuming coupling $\varepsilon$ to be small. To simplify calculations, let us assume parameters
$\gamma, \epsilon \in \Re$; this corresponds to poles lying on axis~$y$.
The equation for the first correction looks like
\begin{align}
-\frac{32 \gamma ^3 \operatorname{e}^{i \alpha' t} \left(x^2+y^2\right) \left(-\gamma ^2+\zeta\right)}{\zeta^3 \left(\epsilon+(x+i y)^2\right)}- \nonumber\\
-\frac{8 \gamma  \operatorname{e}^{-i \alpha' t} \left(\gamma ^2+\left(\epsilon+(x+i y)^2\right) \left(\epsilon-3 (x-i y)^2\right)\right)}{\left(\epsilon+(x-i y)^2\right)^2 \zeta}\nonumber\\
+\frac{8 \gamma ^3 \operatorname{e}^{3 i \alpha' t} \left(\gamma ^2+\left(\epsilon-3 (x+i y)^2\right) \left(\epsilon+(x-i y)^2\right)\right)}{\zeta \left(\epsilon+(x+i y)^2\right)^4}-\nonumber\\
-{\alpha'}^2 \left(1-\frac{3 \gamma ^6+5 \gamma ^4 \zeta+\gamma ^2 \zeta^2}{\zeta^3}\right)u
-\frac{\left(\gamma ^2+\zeta\right)^3}{\zeta^3}\partial^2_t u+\frac{4 i \alpha'  \gamma ^2 \left(\gamma ^2+\zeta\right)^2}{\zeta^3}\partial_t u+\nonumber\\
+\frac{\left(\gamma ^2+\zeta\right)^3 }{\zeta^3}\left(\partial^2_y u+\partial^2_x u\right)
+\frac{8 \gamma ^2 (x+i y) \left(\gamma ^2+\zeta\right)^2}{\zeta^3 \left(\epsilon+(x+i y)^2\right)}\left(\partial_x u+i\partial_y u\right)=0, \label{firsteq}
\end{align}
where
\begin{align}
\zeta = |(x+i y)^2+\epsilon|^2 = \frac{\gamma^2}{u_0u_0^*}.
\end{align}
In the case of distant particles ($\gamma \ll \epsilon$), one can expand the equation in powers of $\frac{\gamma^2}{\zeta}$, which in the leading order leads to
\begin{align}
-\frac{8 \gamma   \operatorname{e}^{-i \alpha' t} \left(\epsilon-3(x-i y)^2\right)}{\left(\epsilon+(x-i y)^2\right)^3}-{\alpha'}^2u-\partial^2_t u+\left(\partial^2_y u+\partial^2_x u\right)=0
\end{align}
and gives asymptotes of the first correction at the spatial infinity, i.e., far from lumps:
\begin{align}
u_1(t,\bar{z})=-\frac{2 i \gamma  \operatorname{e}^{-i \alpha'  t} (2 \alpha'  t-i) \left(\epsilon-3 \bar{z}^2\right)}{{\alpha'} ^2 \left(\epsilon+\bar{z}^2\right)^3}. \label{correction}
\end{align}
It is necessary to notice some peculiarities of the correction found: first, it breaks the analyticity of the solution; secondly, the phase rotates in the opposite direction; and finally, one can see linear growth with time, which means a size increase.
\subsection{Moduli approximation}
Now, we say parameters (moduli) $\gamma= G(t) \operatorname{e}^{i \phi(t) }$, $\epsilon = E(t) \operatorname{e}^{i \theta(t)} $ of ansatz (\ref{nonper}) slowly depend on time. Then, the effective action for such a four-dimensional dynamical system after integrating over $x$, $y$ takes the form [here and below, we denote $\alpha'$ just as $\alpha$; moreover, in the very ansatz, we include rotation with constant velocity $\alpha'$ in the definition of $\phi(t)$]
\begin{align}
&S = \int dt \left( -4\pi +f(E,G)\left(\dot{G}^2+G^2(\dot{\phi}^2-\alpha^2) \right)+\right.\nonumber\\
&\left. + g(E,G) \left(\dot{E}^2+E^2\dot{\theta}^2\right)+ 2h(E,G)\left(\dot{G}\dot{E}+EG\dot{\theta}\dot{\phi}\right) - \frac{\varepsilon}{2}V_{so}\right),
\end{align}

where

\begin{align}
& f(E,G)=\frac{\pi}{2\sqrt{E^2+G^2}}\left(2K(k)-J(k)\right),\label{f}\\
& g(E,G)=\frac{\pi}{2\sqrt{E^2+G^2}}\left(J(k)\right),\label{g}\\
& h(E,G)=\frac{\pi}{2\sqrt{E^2+G^2}}\frac{G}{E}\left(J(k)-K(k)\right),\label{h}\\
& k=\frac{E}{\sqrt{E^2+G^2}},\\
& V_{so}=\int_{0}^{\infty} d\xi \int_{0}^{2\pi} d\psi 32(1-k^2)\xi\left(\frac{\xi^2\cos{(2\psi-\phi)}}{\left(1+\xi^2+2k\xi\cos{(2\psi-\theta)}\right)^2}+\right. \nonumber\\
&\left.+\frac{2k\xi\cos{(\psi+\theta-\phi)}+k^2\cos{(\psi+\phi-2\theta)} }{\left(1+\xi^2+2k\xi\cos{(2\psi-\theta)}\right)^2}\right)^2,
\end{align}

and functions $K(k)$, $J(k)$ are complete elliptical integrals of the first and of the second kinds, respectively. This way, we obtain the explicit form of the motion equations,

\begin{align}
\ddot{G}&= \frac{1}{2\Delta}\left(\frac{\dot{G}^2}{G}(f_4+f_3-f_2)+f_2\left(\frac{2\dot{E}\dot{G}}{E}-\frac{G\dot{E}^2}{E^2}+G\dot{\theta}(\dot{\theta}-2\dot{\phi})\right)+\right.\nonumber\\
&\left.+f_1 G \left(\dot{\phi}^2-\alpha^2\right) \right)+\frac{E^2+G^2}{2\Delta}\left((K(k)-J(k))E G \mathbf{I_2}+J(k)E^2 \mathbf{I_1} \right)\label{Gequation},\\
\ddot{E}&= \frac{1}{2\Delta}\left(\frac{\dot{E}^2}{E}(f_1+f_3-f_2)+f_3\left(\frac{E\dot{G}^2}{G^2}-\frac{2\dot{E}\dot{G}}{G}+E(2\dot{\theta}\dot{\phi}+\alpha^2-\dot{\phi}^2)\right)+\right.\nonumber\\
&\left.+f_4 E \dot{\theta}^2 \right)+\frac{E^2+G^2}{2\Delta}\left((2K(k)-J(k))E^2\mathbf{I_2}+(K(k)-J(k))E G \mathbf{I_1} \right)\label{Eequation},
\end{align}

\begin{align}
\ddot{\theta} &= \frac{1}{\Delta} \left(f_3\left(\frac{\dot{G}}{G}(\dot{\phi}-\dot{\theta})-\frac{\dot{E}\dot{\phi}}{E}\right)-f_4\frac{\dot{E}\dot{\theta}}{E}\right)+\nonumber\\
&+\frac{E^2+G^2}{2\Delta} \left( (K(k)-J(k))\mathbf{I_4}+(2K(k)-J(k))\mathbf{I_3}\right)\label{Tequation},\\
\ddot{\phi}&=\frac{1}{\Delta}\left(f_2\left(\frac{\dot{E}}{E}(\dot{\phi}-\dot{\theta})+\frac{\dot{G}\dot{\theta}}{G}\right)-f_1\frac{\dot{G}\dot{\phi}}{G}\right)+\nonumber\\
&+\frac{E^2+G^2}{2\Delta}\left(J(k) \frac{E^2}{G^2} \mathbf{I_4}+(K(k)-J(k))\mathbf{I_3}\right)\label{Pequation},
\end{align}

where

\begin{align}
& f_1 = J^2(k)\left(E^2+G^2\right)\left(4E^2+G^2\right)- \nonumber\\
&-2J(k)K(k)\left(2E^4+4G^2E^2+G^4 \right)+K^2(k)\left(2E^2+G^2 \right)G^2,\\
& f_2 = 3 J^2(k)\left(E^2+G^2\right)E^2 -2J(k)K(k)\left(E^2+2G^2 \right)E^2+K^2(k)G^2E^2,\\
& f_3 = J^2(k)\left(E^2+G^2\right)G^2 -2J(k)K(k)G^4+K^2(k)G^4,\\
& f_4 = J^2(k)\left(E^2+G^2\right)E^2 -2J(k)K(k)\left(E^2+2G^2\right)E^2+K^2(k)E^2G^2,\\
& \Delta= f_3+f_4,\\
& \mathbf{I_1}=\frac{\varepsilon}{\pi}\sqrt{E^2+G^2} \frac{\partial V_{so}}{\partial G}=\frac{\varepsilon}{\pi}\sqrt{E^2+G^2} \frac{\partial V_{so}}{\partial k}\frac{\partial k}{\partial G},\\
& \mathbf{I_2}=\frac{\varepsilon}{\pi}\sqrt{E^2+G^2} \frac{\partial V_{so}}{\partial E}=\frac{\varepsilon}{\pi}\sqrt{E^2+G^2} \frac{\partial V_{so}}{\partial k}\frac{\partial k}{\partial E},\\
& \mathbf{I_3}=\frac{\varepsilon}{\pi}\sqrt{E^2+G^2} \frac{\partial V_{so}}{\partial \theta},\\
& \mathbf{I_4}=\frac{\varepsilon}{\pi}\sqrt{E^2+G^2} \frac{\partial V_{so}}{\partial \phi}.
\end{align}

Equations (\ref{Gequation})--(\ref{Pequation}) can be solved numerically. Then, $\gamma(0), \epsilon(0)$ give the initial configuration,
$\dot{\gamma(0)}$, $\dot{\epsilon(0)}$ -- a small perturbation. In our case, $\dot{G},\dot{E},\dot{\theta}=0$, $\dot{\phi}=\alpha$.
Equations were solved using Runge-Kutta fourth-order method for the following parameter values: couplings $\alpha =0.03$, $\varepsilon = 0.1$, lumps' size $\frac{G}{2\sqrt{E}}=1$,
and the distance between lumps $2 \sqrt{E}=50$. Integrals $\mathbf{I_1}$--$\mathbf{I_4}$ were also calculated numerically.

\section{Results}

Several effects of spin-orbit interaction on the two-particle solution were discovered:

\begin{enumerate}

\item Size increase: this effect is in good correspondence with the growth found with perturbation theory.
\begin{figure}[h]
\center
\includegraphics[scale=0.6]{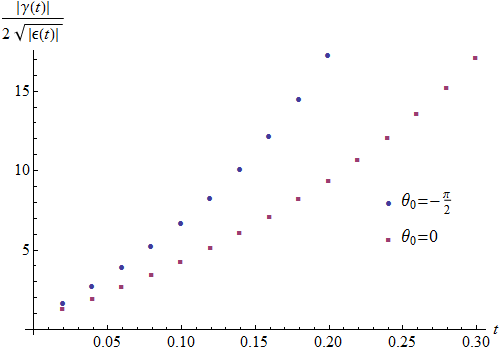}
\caption{Particle size evolution}
\label{size}
\end{figure}

\item Attraction: the distance between particles declines, which was not seen from correction (\ref{correction}). The fact, however, is because expression
(\ref{correction}) gives only asymptotes at spatial infinity, which remain the same.
\begin{figure}[h]
\center
\includegraphics[scale=0.6]{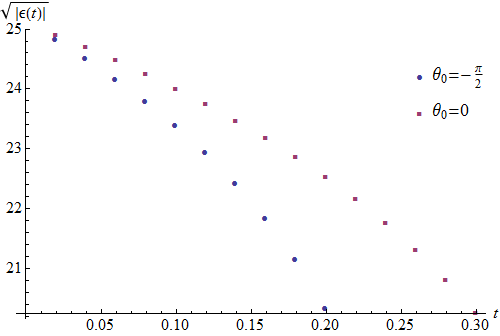}
\caption{Evolution of distance between particles}
\label{distance}
\end{figure}

\item Dependence on phases: figures~\ref{size} and \ref{distance} show numerical solutions for two choices of parameters $\theta_0=-\frac{\pi}{2}$ and $\theta_0=0$
that determine the initial positions of the Q-lumps. The first one corresponds to particles lying on the real axis, and the second one corresponds to particles lying on the imaginary one. While system behavior remains qualitatively the same, one can notice that the velocities of processes depend significantly on the parameter. Thus, in contrast to functions (\ref{f})--(\ref{h}) not depending on phases $\theta$ and $\phi$ and allowing the consideration of just functions $G(t)$ and $E(t)$ in the absence of spin-orbit interaction [in the case of the same initial $\dot{\theta}(0)=0$ and $\dot{\phi}(0)=\alpha$], the addition of $\mathcal{L}_{so}$ results in the appearance of such a dependence.
In Fig.~\ref{profilepicture}, one can see how potential $V_{so}(k)$ depends on phases: indeed, for every fixed $\theta$, $\phi$ profile, $V_{so}(k)$
qualitatively remains the same and results in the growth and attraction of particles.
\begin{figure}[h]
\center
\includegraphics[scale=0.45]{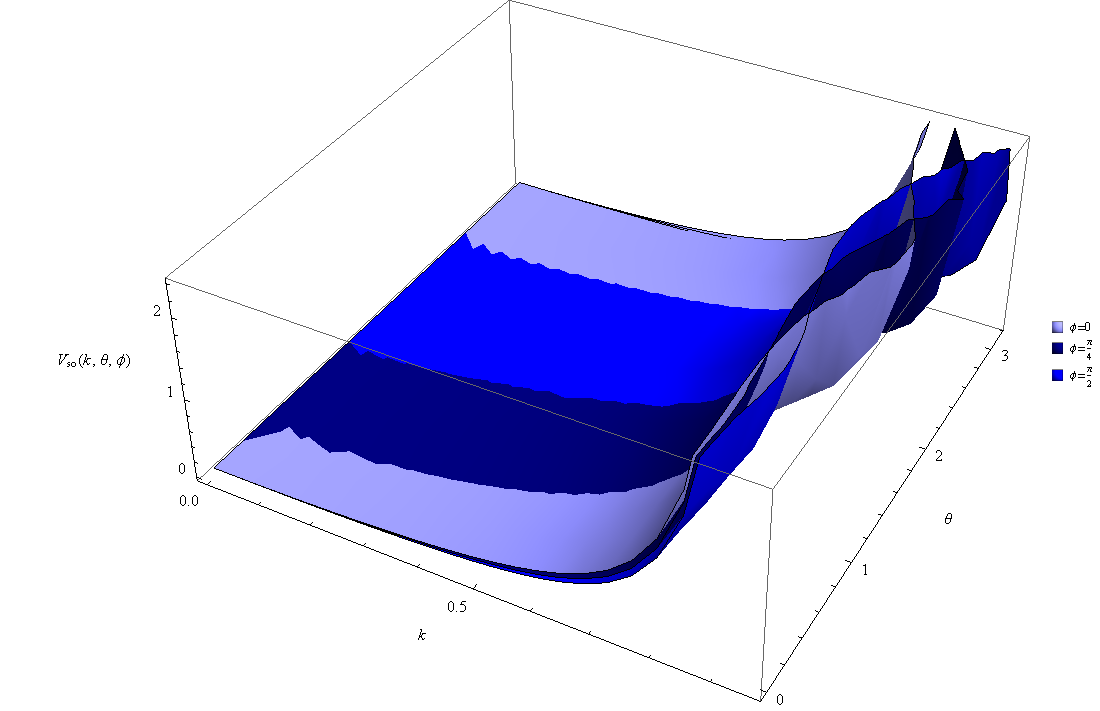}
\caption{Spin-orbit interaction potential $V_{so}(k,\theta,\phi)$}
\label{profilepicture}
\end{figure}

\end{enumerate}

\section{Conclusion}

In this paper, we considered the O(3)~$\sigma$ model with an additional potential term on the domain wall in 3+1-dimensional spacetime admitting solitonic solutions (Q-lumps). Effects caused by spin-orbit interaction in the bulk on two-particle solutions were studied.

The asymptotic behavior at spatial infinity of the first corrections to the two-particle ansatz was found perturbatively; the evolution of the initially motionless configuration was numerically studied through adiabatic approximation. We discovered that in the presence of spin-orbit interaction Q-lumps begin to interact: they grow and attract; also, entanglement between the internal degrees of freedom and coordinates leads to dependence on phases of the configuration.

One important feature of Q-lumps in (2+1) dimensions discovered by Leese is that the additional term, which significantly changes the behavior of the system, can be treated as just external perturbation. The interaction effects outlined in Sec.~4 suggest that another perturbation to the action in the media (in the bulk) can "break" some properties of Leese's noninteracting lumps living on the (2+1)-dimensional wall, and all this happens even when taking both small couplings.
\section*{Acknowledgments}
The author is grateful to A.~Gorsky for suggesting the problem and for numerous inspirational discussions and also would like to thank K.~Pavlenko for valuable comments. The work was supported in part by Grant No. RFBR-16-02-00252.

\end{document}